\documentclass[A4,12pt,oneside]{article}
\usepackage{amssymb, amscd, amsmath,amsthm,amsfonts,enumerate}
\usepackage{graphicx}
\usepackage[utf8]{inputenc}
\usepackage{graphics}
\usepackage{epsfig}
\usepackage{placeins}
\usepackage{color}
\usepackage{amsfonts}
\usepackage{xcolor}
\usepackage{booktabs}
\usepackage{float}
\usepackage{subfigure}
\usepackage{comment}
\usepackage{geometry}
\usepackage{lscape}
\usepackage{tikz}
\usetikzlibrary{calc}
\usepackage{cite}
\usepackage{caption}
\usepackage{blindtext}
\usepackage{nomencl}
\usepackage{commath}
\usepackage{float}

\usepackage{hyperref}

\usepackage[english]{babel}

\begin{document}

\begin{center}
	{\Large \bf A Comparative Predicting Stock Prices using Heston and Geometric Brownian Motion Models}\\[12mm]
	
	{\bf  	H. T. Shehzad\footnote{Department of Mathematics, School of Science and Engineering, Lahore University of Management Sciences,	Opposite Sector U, DHA, Lahore Cantt., 54792, Pakistan, 19070012@lums.edu.pk},  M. A. Anwar\footnote{Department of Mathematics, School of Science and Engineering, Lahore University of Management Sciences,	Opposite Sector U, DHA, Lahore Cantt., 54792, Pakistan}, 
  M. Razzaq\footnote{Weierstrass Institute, Mohrenstrasse 39, 10117, Berlin, Germany, m.razzaq@ianus-simulation.de, mrazzaq@math.tu-dortmund.de} }
\end{center}

 \section*{ Abstract}
This paper presents a novel approach to predicting stock prices using technical analysis. By utilizing Ito's lemma and Euler-Maruyama methods, the researchers develop Heston and Geometric Brownian Motion models that take into account volatility, interest rate, and historical stock prices to generate predictions. The results of the study demonstrate that these models are effective in accurately predicting stock prices and outperform commonly used statistical indicators. The authors conclude that this technical analysis-based method offers a promising solution for stock market prediction.

\textbf{ Keywords}: Euler-Maruyama; Ito’s lemma; Numerical methods;  Pakistan Stock Exchange; Comparative Study; Stochastic Differential Equation.

\section{Introduction}

The Heston and Geometric Brownian Motion (GBM) models are two common models used to predict stock prices.

The Heston model is a stochastic volatility model that takes into account both the level of volatility and the correlation between stock price and volatility. It assumes that the stock price and volatility follow a bivariate stochastic process, allowing for the modeling of volatility clustering, which is a common phenomenon in financial markets. The Heston model is often used in options pricing and risk management.

On the other hand, the Geometric Brownian Motion model is a simple and widely used model for stock price simulation. It assumes that stock prices follow a random walk, with the log returns being normally distributed. The model also assumes constant volatility, which is a key simplifying assumption. This model is often used as a benchmark for comparing the performance of more complex models, such as the Heston model.

In terms of predicting stock prices, the Heston model may provide more accurate results because it takes into account the volatility and correlation between stock price and volatility. However, the complexity of the Heston model may also make it more difficult to implement and calibrate compared to the GBM model.

Ultimately, the choice between the Heston and GBM models depends on the specific needs of the user and the type of analysis they are conducting. If a simple and quick prediction is sufficient, the GBM model may be the way to go. If a more accurate prediction is needed, taking into account the volatility and correlation between stock price and volatility, the Heston model may be a better choice.

One of the fastest-growing areas of mathematical computing is the financial industry, with the emergence of disciplines such as financial mathematics, financial engineering, and computational finance \cite{1}. The growth of this area can be attributed to the challenging intellectual problems underlying financial markets, which have attracted researchers from different disciplines. In this context, we will discuss a numerical challenge in financial mathematics, which is asset pricing \cite{3}.

There have been numerous studies published on asset pricing over the years, using various methodologies, including the Black-Scholes model \cite{4}, jump-diffusion model \cite{5}, Heston model \cite{6}, Binomial tree model \cite{7}, and Monte Carlo methods \cite{8}. All of these models except the Binomial tree model make use of Stochastic Differential Equations (SDEs) to overcome the challenges in asset pricing.

Recently, researchers have used uncertain differential equations in asset pricing, such as Yuan Liu \cite{9} who built an uncertain currency model to derive European and American market option pricing. R. Gao \cite{10} and Xin Gao \cite{11} both developed models that were defined by unknown differential equations involved in a Liu mechanism that changes over time in a set of uncertain variables, primarily examining the pricing of the American and Asian barrier options in the volatile stock market. Shican Liu \cite{12} extended the Heston model to multi-scale stochastic volatility and jump-diffusion-based hybrid option pricing model, taking into account the correlation results. Zhe Li \cite{13} evaluated the value of liquidity risk in the discrete barrier option pricing and suggested a new model explaining asset price dynamics in the presence of jumps and liquidity risks.

Quantum computers are also being increasingly used in option pricing. Patrick Rebentrost \cite{13} introduced a quantum algorithm for Monte Carlo pricing of financial instruments, which shows how the related probability distributions can be planned in a quantum superposition and how the price of financial derivatives can be derived through quantum measurements. Nikitas Stamatopoulos \cite{14} suggested a strategy for pricing options and investments of options on a gate-based quantum computer, using an amplitude measurement algorithm that provides quadratic speed compared to traditional Monte Carlo techniques.

Provision, as defined by \cite{2}, involves evaluating past data to make forecasts about future events, encompassing a wide range of subjects including business, industry, economics, ecology, and finance. Forecasting difficulties can be categorized as short-term, medium-term, and long-term, and are closely tied to the temporal analysis of data. A time series analysis of a particular variable, such as a uniform stock price, can be used to examine the maturational chain of events. Uni-variate data, which only describes one specific stock, can be improved with time series analysis to uncover patterns, trends, periods, or cycles.

For example, in the stock market, an early understanding of whether the market will rise (bullish) or fall (bearish) can greatly impact investment decisions. Time series analysis can also help identify the most promising firms at a specific point in time. Given its importance, time series analysis and forecasting are highly valued areas of study \cite{3}.

Provision, which is the process of evaluating past data to forecast future events \cite{2}, covers a wide range of subjects, including business and industry, economics, ecology, and finance. One of the main challenges with prediction is the difficulty in determining the temporal scale of the forecast, which can be characterized as short-term, medium-term, or long-term \cite{2}. In the case of stock prices, this difficulty is addressed through time series analysis, which helps in identifying patterns, trends, periods, and cycles in the data.

Two key techniques are used for stock price prediction: fundamental analysis and technical analysis \cite{2}. Fundamental analysis is an investment study that evaluates a stock's value by examining the company's sales, profits, and other economic elements. This technique is considered the best approach for long-term predictions. On the other hand, technical analysis uses historical stock prices to identify future prices. One commonly used technique in technical analysis is average movement analysis.

Additionally, to perform asset pricing, various financial models are used such as Capital Asset Pricing Model (CAPM) and Arbitrage Pricing Theory (APT). These models consider various factors, including risk, return, and cost of capital, to determine the fair value of an asset \cite{1}.

Moreover, the efficient market hypothesis (EMH) is another important concept in asset pricing. EMH posits that it is impossible to consistently achieve returns that are above average by using any information that is publicly available \cite{4}. This means that the current market price of an asset reflects all relevant information and it is impossible to consistently buy low and sell high.

In conclusion, asset pricing is a crucial aspect of financial management as it helps investors make informed decisions and evaluate the potential return on their investments. A combination of fundamental analysis, technical analysis, financial models, and market hypothesis helps in determining the fair value of an asset and predicting future stock prices.

In recent years, there has been a shift towards using Stochastic Differential Equations (SDEs) for stock price prediction \cite{1}. The advancement in financial mathematics and stochastic diffusion models has allowed for the solution of numerous complex asset pricing problems. One of the most well-known and reliable models for asset pricing is the Black-Scholes-Merton (BSM) model \cite{1}. In the BSM model, the Geometric Brownian Motion (GBM) with constant growth rate and standard deviation is used to model asset pricing in financial mathematics \cite{1}.

Traders and investors often use past information, such as previous stock prices, to predict asset prices and market movements, and make more informed investment decisions. However, this type of information is not accounted for in classical SDEs. To address this issue, delay terms, such as drift, volatility, and diffusion coefficients, are included in the GBM model \cite{8}. This allows for the consideration of previous information and a response to be included in the model, making it more efficient and effective in predicting stock prices.

This paper is structured as follows. Section \ref{Math} provides the mathematical formulation of the models used. In subsection \ref{heston}, the original work of Heston (1993) is adapted to the context of the Pakistan Stock Exchange (PSX), as the model is highly effective in capturing the behavior of equities in the PSX market. The subsection \ref{gbm} discusses the GBM model and its parameters. The Ito lemma, which serves as the foundation of the GBM model, is utilized to solve stochastic differential equations. Finally, in Section \ref{eng}, the models are applied to data from the PSX to forecast future stock prices for Engro Fertilizers Limited.

\section{Mathematical Formulation} \label{Math}
We will begin by discussing the Heston Model, used in finance to price options while considering both volatility and correlation. We will use the Euler-Maruyama technique to solve the model by breaking it down into smaller intervals and using random walks. Then, using Ito's Lemma, we will solve a linear stochastic differential equation that represents the Geometric Brownian Motion model, which describes stock prices as a random walk with drift and diffusion terms, commonly used for option pricing.

\subsection{Heston Model} \label{heston}
The Heston Model is a financial model defined by two 
Stochastic Differential Equations (SDEs) (for details see \cite{11,12,13})
\begin{equation}
    dX_t= \mu X_t dt+\sqrt{\zeta_t}X_t dW_{t_1}
\end{equation}
\begin{equation}
    d\zeta_t = \kappa(\theta-\zeta_t )dt+\sigma_\zeta \sqrt{\zeta_t}dW_{t_2}
\end{equation}
the equations are expressed in terms of Wiener processes, represented by $W_{t_1}$ and $W_{t_2}$, and a correlation term $\rho$ given by 
\begin{equation}
    dW_{t_1} dW_{t_2} =\rho dt.
\end{equation}

By applying the Euler-Maruyama method \cite{14} to the Heston Model, the resulting discrete solution is obtained
$$X_t = X_{t-1} +\mu X_{t-1}dt+\sqrt{\zeta_{t-1}}X_{t-1}\sqrt{dt}Z_{t_1} $$
$$\zeta_t = \zeta_{t-1} +\kappa(\eta- \zeta_{t-1})dt+\sigma_\zeta \sqrt{\zeta_{t-1}dt}Z_{t_2} $$
$$Z_{t_1}= G_{t_1}$$
$$Z_{t_2} = \rho G_{t_1} +\sqrt{1-\rho^2}G_{t_2},$$
%
the solution is expressed in terms of independent, identically distributed random variables $G_{t_1}$ and $G_{t_2}$, or i.i.d. random variables.
\subsection{Geometric Brownian Motion Model (GBM)} \label{gbm}
The Geometric Brownian Motion Model (GBM) is a type of stochastic function that is differentiable and operates based on GBM \cite{6,7,8} 
\begin{equation}
     dX_t=\mu (X_t,t)dt+\sigma(X_t,t)dW_t.  
\end{equation}

By utilizing Taylor's theorem 
\begin{align*}
    &g(X_t+dX_t)=g(X_t)+dX_tg'(X_t)+\frac{(dX_t)^2}{2}g''(X_t)+\cdot\cdot\cdot
    \\&+\frac{(dX_t)^{(n-1)}}{(n-1)!}g^{(n-1)}(X_t)+\frac{(dX_t)^n}{n!}g^{(n)}(X_t).
    \end{align*} 
If $X_t$ is a non-stochastic variable, and if we set $dX_t\longmapsto 0$, then the following can be derived:
\begin{equation}
    g(X_t+dX_t)-g(X_t)=g'(X_t)dX_t=d(X_t)
\end{equation}
however, the term $(dX_t)^2$ is important in stochastic processes. Hence, we let $X_t$ be a stochastic variable and suppose that the stochastic process $X_t$ is represented by equation $(4)$
\begin{equation}
    (dX_t)^2=\mu^2(dt)^2+\sigma^2(dw_t)^2+2\mu\sigma dtdw_t =  \sigma^2(dw_t)^2=\sigma^2dt. 
\end{equation}

Now, if we consider the case of a function $g(X_t, t)$, where $X_t$ is a stochastic process, we can extend the above equation to get

\begin{equation}
dg(X_t,t)=\frac{\partial g}{\partial X_t}dX_t+\frac{\partial g}{\partial t}dt+\frac{1}{2} \biggl({\frac{\partial^2 g}{\partial X_t^2}(dX_t)^2+\frac{2\partial^2g}{\partial X_t \partial t }dX_tdt+\frac{\partial^2 g}{\partial t^2}dt^2}\biggr).
\end{equation}
The first term on the right-hand side of the equation represents the change in the function $g$ due to the change in the process $X_t$, while the second term represents the change in the function $g$ due to the change in time $t$. The third term on the right-hand side of the equation takes into account the effect of the change in the process $X_t$ and time $t$ on the change in the function $g$, which includes the second-order derivatives of the function $g$ with respect to $X_t$ and $t$. 
The function $g(X_t, t)$, where $X_t$ is a stochastic process, can be extended using Taylor's theorem by neglecting higher-order derivatives. This results in the equation 
\begin{equation}
dg(X_t,t) = \frac{\partial g}{\partial X_t}dX_t + \frac{\partial g}{\partial t}dt + \frac{1}{2} \biggl({\frac{\partial^2 g}{\partial X_t^2}dX^2 + \frac{2\partial^2g}{\partial X_t \partial t }dX_tdt + \frac{\partial^2 g}{\partial t^2}dt^2}\biggr). 
\end{equation}
%
%
The equation from stochastic calculus can be simplified by taking into account that $(dt)^2 = 0$ and $dX_tdt = 0$. This results in the following simplified expression:
\begin{equation}
dg(X_t,t) = \frac{\partial g}{\partial X_t}dX_t + \frac{\partial g}{\partial t}dt + \frac{1}{2}\frac{\partial^2 g}{\partial X_t^2}(dX_t)^2 
\end{equation}
by incorporating the results discussed above into equation $(3.6)$, we get the following expression:
\begin{equation}
dg(X_t,t)=\biggl(\mu \frac{\partial g}{\partial X_t}+\frac{\partial g}{\partial t} + \frac{1}{2}\sigma^2\frac{\partial^2 g}{\partial X_t^2}\biggr)dt + \sigma \frac{\partial g}{\partial X_t}dw_t, 
\end{equation}
this result is commonly referred to as Ito's lemma, as described in reference \cite{9}.

\textbf{Solution of  Stochastic Differential Equation (SDE)}

The equation $(3.7)$ is a Stochastic Differential Equation (SDE) \cite{12} that describes the evolution of a random variable $X_t$. The rate of return is represented by the parameter $\mu(X_t,t)$ and the volatility is represented by $\sigma(X_t,t) = \sigma X_t$. The SDE is given by the equation 
\begin{equation}
 dX_t = \mu(X_t,t)dt + \sigma(X_t,t)dw_t,
 \end{equation}
 where $dw_t$ represents the Wiener process. This equation captures the uncertainty in the evolution of $X_t$ due to the presence of randomness.  By integrating both sides of the equation, we get:
 \begin{equation}
 \int_{0}^{t}\frac{dX_s}{X_s}= \int_{0}^{t}\mu ds+ \int_{0}^{t}\sigma dW_s.  
\end{equation}
The first integral in the above equation, which contains no random terms, has a constant coefficient $\mu$ and results in $$\int_{0}^{t}\mu ds = \mu t.$$ The second integral has a coefficient of $dW_s$, an invariant period time constant, which results in $$\int_{0}^{t}\sigma dW_s = \sigma[W_t - W_0],$$ where $W_0 = 0$. This leads to the final result of $$X_t = X_0 e^{(\mu-\sigma^2/2)+\sigma W_t}.$$

By using Ito's lemma, we arrive at the following expression for the derivative of $X_t$:
$$dX_t = X_0 e^{(\mu-\frac{1}{2}\sigma^2)t + \sigma W_t}\left((\mu - \frac{1}{2}\sigma^2)dt + \sigma dW_t + \frac{1}{2}\sigma^2 dt\right)$$
 %
\section{Results} \label{eng}
Engro Fertilizers Limited is a subsidiary of Engro Corporation listed on all 3 stock exchanges in Pakistan with an 86\% stake in the production and marketing of urea and NPK fertilizers. Engro Urea, Pakistan's well-known household urea brand, is the country's leading player in terms of capacity, with diverse ventures contributing to the improvement of its production capacity from 173 kt/year to 975 kt/year.

To predict the future stock prices of Engro Fertilizers, we employed the Heston model and utilized the Euler Maruyama process for its solution. The results were close to the actual prices but with the possibility of being even better. Unfortunately, we were unable to estimate the prices of call and put options as options trading has not been introduced in Pakistan and there is a lack of data for the parameters required for the forecasting of call and put options. We also used the GBM model for price prediction comparison with the Heston model and found that the latter produced more accurate values. The data was collected from EFERT on March 13, 2021, from the Pakistan Stock Exchange (PSX).

\begin{table}[ht]
\centering
\caption{Comparison of Heston and GBM Models in Forecasting Engro Fertilizers Limited Stock Price}
\begin{tabular}[t]{p{5cm} p{5cm} p{5cm}}
\hline
&Heston Model&GBM Model\\
\hline
Variation of Prices&1 day&1 day\\
Simulated Range&$67.2098 - 68.2224$&$67.2987 - 68.1559$\\
Exact Range&$67.20 - 68.22$&$67.20 - 68.22$\\
\hline
\end{tabular}
\label{t1}
\end{table}%
In conclusion, table \ref{t1} provides a comparison between the stock price prediction of Engro Fertilizers Limited using the Heston Model and the GBM Model. Both models were used to simulate the stock price for a period of 1 day with the same input parameters. The results show that the simulated range of the Heston Model is $67.2098 - 68.2224$, while the GBM Model has a simulated range of $67.2987 - 68.1559$. The exact range of the stock price is $67.20 - 68.22$ for both models. The results suggest that the Heston Model provides a more accurate prediction of the stock price compared to the GBM Model.

\begin{table}[ht]
\centering
\caption{Error Calculations.}
\begin{tabular}[t]{p{5cm} p{5cm} p{5cm}}
\hline
Absolute Error&Heston Model&GBM Model\\
\hline
Minimum&0.0098&0.0987\\
Maximum&0.0024&0.0641\\
\hline
\end{tabular}
\label{t2}
\end{table}%
Table \ref{t2} shows the error calculations for the stock price prediction of Engro Fertilizers Limited. The minimum and maximum absolute error values are shown for both the Heston Model and the GBM Model. As can be seen from the table, the Heston Model has a smaller minimum and maximum absolute error compared to the GBM Model. This suggests that the Heston Model provides a more accurate prediction of the stock price compared to the GBM Model.
\FloatBarrier



\begin{figure}
     \begin{subfigure}
     \centering
     \includegraphics[width=7in]{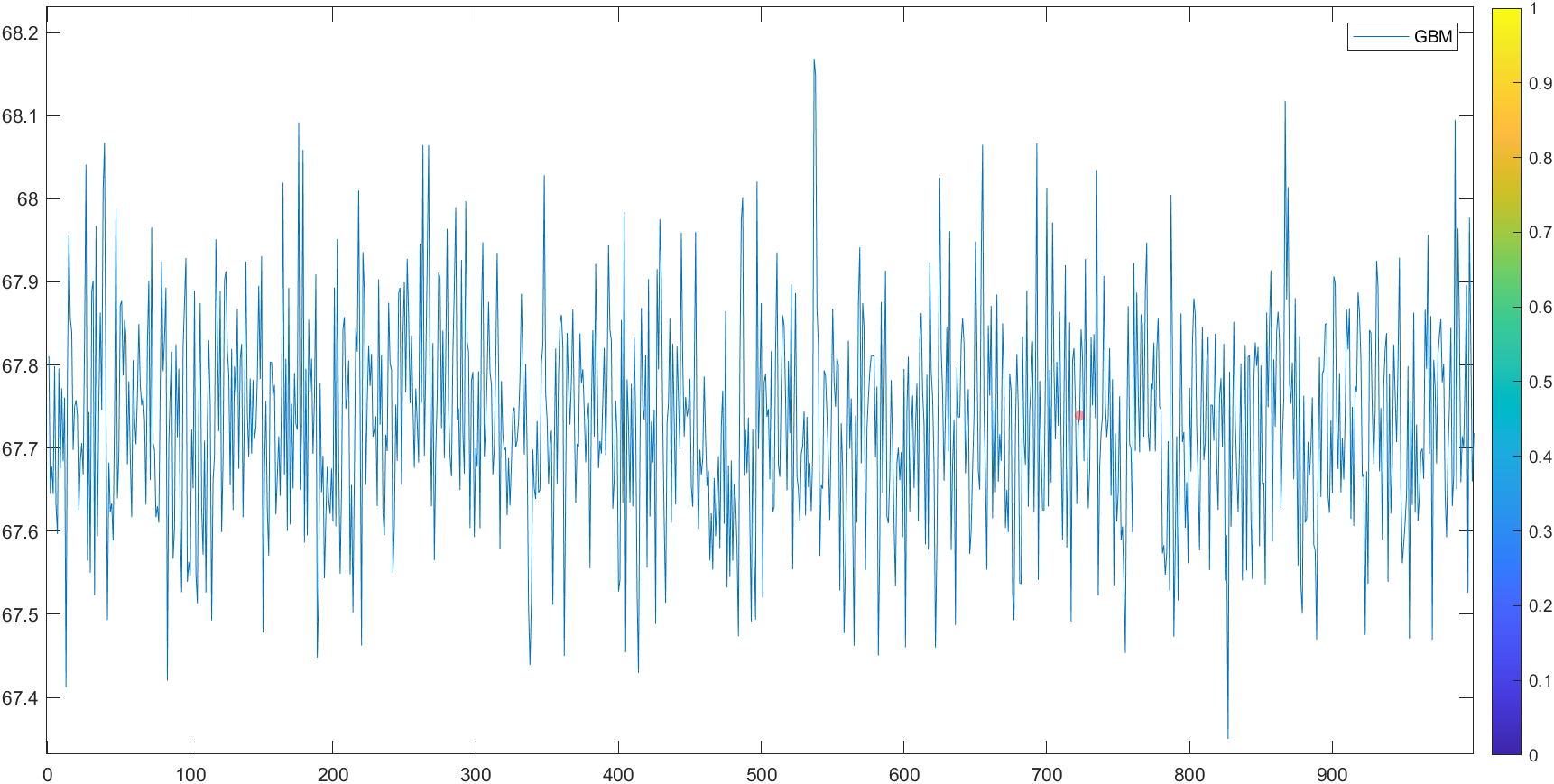}
     \caption*{(a): Movement of stock prices over 10000 simulations by GBM, time (1 Day)}
     \end{subfigure}
     \begin{subfigure}
     \centering
     \includegraphics[width=7in]{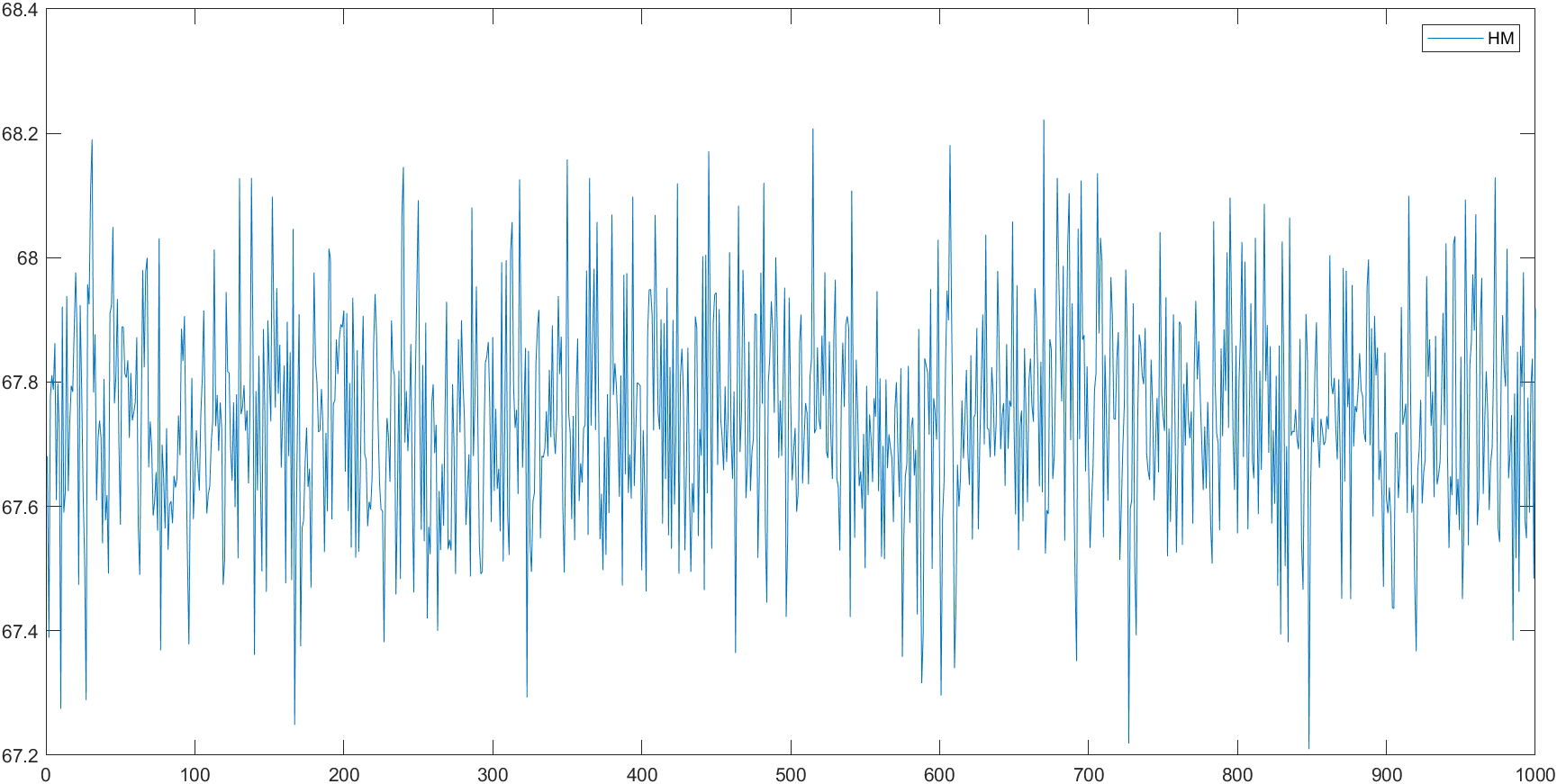}
     \caption*{(b): Movement of stock prices over 10000 simulations by Heston model, time (1 Day)}
     \end{subfigure}
\caption{Comparison of GBM and Heston models}
    \label{fig:example}
\end{figure}

\FloatBarrier
The above figure \ref{fig:example} presents a comparison of the movement of Engro Fertilizers Limited's stock prices over 10000 simulations for a time period of 1 day. The first subfigure (a) illustrates the movement of stock prices by using the Geometric Brownian Motion (GBM) model. The second subfigure (b) shows the movement of stock prices using the Heston model. These figures provide a visual representation of the predictions made by the two models and can be used to compare their performance.

The stock price of EFERT is represented on the y-axis, while the simulation is shown on the x-axis. The correlation parameter $\rho$ is equal to 0.00165, the long run variance $\theta$ is equal to -0.09228, the mean reversion rate $\kappa$ is equal to 0.00979, the time $T$ is equal to 1/252, which is equal to 1, the volatility $\sigma$ is equal to 0.03, and the rate of return $r$ is equal to 0.513.

The research paper on stock price prediction for companies listed on the Pakistan Stock Exchange is significant for its ability to provide useful information for investment decisions, improve market efficiency, and advance academic knowledge in financial modeling. The use of Heston and GBM models in the analysis helps to manage risks and increase transparency in the financial markets. The findings of the study have the potential to impact the country's economy and drive economic growth by providing valuable insights to business leaders and policymakers.

\section{Conclusions}
In conclusion, the study conducted a survey of Heston and GBM models for their validity of parameters such as mean reversion rate, long-run variance, correlation parameters, etc. The algorithms were implemented to estimate EFERT stock prices in PSX and their time variations, with Monte Carlo and Euler Maruyama techniques used for simulation. The results showed that the Heston model gave better results compared to the GBM model, with more accurate results to two decimal places. These models can be further extended for the prediction of other volatile markets, such as future and oil prices, based on available data sets.

The focus of this paper is on the application of stochastic differential equations in finance. We examine the validity of parameters such as mean reversion rate $\kappa$, long-run variance $\theta$, and correlation parameter $\rho$ in the Heston and GBM models through a literature survey. We use the algorithms from \cite{1} to implement the Heston and GBM models and apply them to estimate the stock prices of EFERT on the PSX and their time variations. The Monte Carlo and Euler Maruyama techniques are employed to simulate the models, and the results are presented in graphical and tabular form. 
Our findings indicate that:
\begin{itemize}
    \item The Heston model produces more accurate results than the GBM model.
    \item The Heston model has an accuracy of up to two decimal places.
    \item The GBM model has an accuracy of up to one decimal place.
\end{itemize} 
These models can be further extended for predicting future prices of commodities such as oil and others, considering the high volatility of these markets.

\section*{Acknowledgments}
The authors did not receive any financial support for conducting this research or publishing this paper.

\end{document}